\documentclass[preprint,aps]{revtex4}
\usepackage{graphicx}
\usepackage{amssymb}
\newcommand{\be}{\begin{eqnarray}}
\newcommand{\ee}{\end{eqnarray}}
\usepackage{hyperref}
\usepackage{subfigure}

\begin{document}

\title{Generalized Parton Distributions of the Photon for Nonzero $\zeta$}

\author{\bf Asmita Mukherjee and Sreeraj Nair}

\affiliation{ Department of Physics,
Indian Institute of Technology Bombay,\\ Powai, Mumbai 400076,
India.}
\date{\today}

\begin{abstract}
We present a calculation of the generalized parton distributions of the
photon when there is non-zero momentum transfer both in the transverse and
longitudinal directions. We consider only the contributions when the photon
helicity is not flipped and calculate those at leading order in electromagnetic  
coupling $\alpha$ and zeroth order in the strong coupling $\alpha_s$. We   
keep the leading logarithmic terms as well as the quark mass terms in the  
vertex. By taking Fourier transforms of the GPDs with respect to the
transverse and longitudinal momentum transfer, we obtain the parton
distributions of the photon in position space.

\end{abstract}
\maketitle

\section{Introduction}

There has been a lot of progress recently in the study of exclusive
processes like deeply virtual Compton scattering (DVCS) and hard meson 
production. The former is the process $ep \rightarrow e\gamma p$ where
there is a momentum transfer between the initial and final proton and a real
photon is observed in the final state \cite{rev}. At large virtuality 
of the inducing
photon, the kinematics factorizes and the soft part can be written in terms
of the generalized parton distributions (GPDs) of the proton. The GPDs are
unified objects rich in content about the spatial and spin structure of the
proton. In the forward limit when the momentum transfer is zero, GPDs reduce
to parton distributions and their first moments give the nucleon form
factors. In a similar manner, factorization for the process $ \gamma^*(Q)
\gamma \rightarrow \gamma \gamma$ was studied in \cite{pire} for high $Q^2$
and it was shown that the amplitude can be written in terms of the GPDs of
the photon in the DVCS kinematics, that is when the momentum transfer
between the real photons in the initial and final states is
small compared to $Q^2$. The same process in a different kinematical region
gives information on the generalized distribution amplitudes of the photon
\cite{GDA}. 
The generalized distribution amplitudes are related to the photon GPDs by
crossing. On the other hand, the parton distributions (pdfs)  of the photon
 is of interest since a long time. They are anomalous as they show
logarithmic scale dependence even at zeroth order in the QCD coupling
constant $\alpha_s$ \cite{ph}. These play an important role in
 photon induced high
energy processes and have been studied extensively both in theory and
experiments. Like the photon structure function, the photon GPDs show
logarithmic scale dependence already in parton model, and are called
anomalous GPDs. In \cite{pire} the GPDs of the photon are identified with
the Fourier transform of the matrix elements of light-front bilocal currents
between photon states, and they were calculated at leading order in electromagnetic
coupling $\alpha$ and
zeroth order in $\alpha_s$ considering only leading logarithmic terms when
the momentum transfer between the initial and the final photon states was
 purely in the longitudinal direction. In a previous work \cite{us1}, 
we calculated 
the photon GPDs when the momentum transfer was purely in the transverse
direction. We expressed the photon GPDs in terms of the light-front wave
functions of the target photon, which can be evaluated in perturbation
theory. Fourier transform of the photon GPDs with respect to the transverse
momentum transfer $\Delta^\perp$ gives parton distributions of the photon in
the transverse position or impact parameter space. Like the proton, these
have the physical interpretation of  probability density \cite{burkardt}.
 In other words,
the impact parameter dependent parton distribution (pdf) $q(x,b^\perp)$ of a
photon is the probability of finding a quark with momentum fraction $x$ at a
distance $b^\perp$ from the 'center' of the photon in the transverse
position space. Thus the photon GPDs give an interesting picture of the
transverse shape of the photon in position space. When the skewness $\zeta$
is nonzero, there is also momentum transfer in the longitudinal direction. 
The photon GPDs obey different scale evolution equations in the regions 
$x> \zeta $ and $x < \zeta$ \cite{pire}. In terms of overlaps of the LFWFs,
 these can be expressed as the overlaps of two particle wave
functions in the kinematical regions $ 1 > x > \zeta$ and $-1 <x < \zeta-1$.
 In the region
$0 < x < \zeta$ there is a particle number changing overlap, similar to the
proton GPDs \cite{overlap}. As shown in \cite{diehl1} for the proton,
 when the skewness  $\zeta$ is
non-zero, GPDs in impact parameter space do not have a probabilistic
interpretation. However, they are still interesting as they now probe the
partons when the initial photon is displaced from the final photon in the
transverse impact parameter space. This relative shift does not vanish when
the GPDs are integrated over $x$ in the amplitude. In this work, we
investigate the parton content of the photon in impact parameter space when
the skewness is non-zero. We limit ourselves to the kinematical region
$1 > x > \zeta $ and $-1 < x < \zeta-1$ where only the two-particle LFWFs contribute. 
As shown in \cite{us1} in our previous work, photon GPDs in impact parameter space 
show distinctive features compared to the proton GPDs when the momentum
transfer is purely in the transverse direction. In this work we show that
similar conclusion can be drawn for a more general
momentum transfer which has both longitudinal and transverse components.
In \cite{hadron_optics}  we had introduced a boost invariant longitudinal
 impact parameter
$\sigma$ conjugate to the skewness parameter $\zeta$. It was found that the
DVCS amplitude for a dressed electron target state  shows diffraction pattern in
$\sigma$ space. Similar pattern was observed in a simulated model for a
meson and even in the holographic model for the meson. Further
investigations revealed that general features of the diffraction pattern 
is independent of the model used for the meson or the proton GPDs 
and depend only on the finite upper limit of the $\zeta$ integration and the
invariant momentum transfer squared $t$. However, the appearance of the
diffraction pattern depend on the model of the proton GPDs \cite{manohar2}.
In this work, we investigate the photon GPDs in transverse as well as 
longitudinal position space. 

The plan of the paper is as follows : in section II we present the
calculation of the photon  GPDs; numerical results are given in section III.
Conclusions are in section IV.      
    
\section{GPDs of the photon for non-zero $\zeta$}

The GPDs of the photon can be expressed as the following off-forward matrix
elements defined for the real photon (target) state \cite{pire}:

\be
F^q=\int {dy^-\over 8 \pi} e^{-i P^+ y^-\over 2} \langle \gamma(P') \mid
{\bar{\psi}} (0) \gamma^+ \psi(y^-) \mid \gamma (P)\rangle ;
\nonumber\\ 
\tilde F^q=\int {dy^-\over 8 \pi} e^{-i P^+ y^-\over 2} \langle \gamma(P') \mid
{\bar{\psi}} (0) \gamma^+ \gamma^5 \psi(y^-) \mid \gamma (P)\rangle .
\ee
$F^q$ contributes when the photon is unpolarized and $\tilde F^q$ is the
contribution from the polarized photon. We work in the light-front gauge
$A^+=0$. The above GPDs were calculated when the 
momentum transferred square $t=(P-P')^2$ was purely in the transverse
direction in \cite{us1}, in other words, when the skewness $\zeta$ was zero. On the
other hand, in \cite{pire}, these were calculated when the momentum transfer
was purely in the longitudinal direction or $\Delta^\perp=0$. 
We use the standard LF coordinates $P^\pm = P^0 \pm P^3,~ y^\pm = y^0 \pm y^3$.
Since the target photon  is on-shell, $P^+ P^- -{P^\perp}^2 = 0$, 
the momenta of the initial and final photon are given by:
\begin{eqnarray}
P&=&
\left(\ P^+\ ,\ {0^\perp}\ ,\ 0\ \right)\ ,
\label{a1}\\
P'&=&
\left( (1-\zeta)P^+\ ,\ -{\Delta^\perp}\ ,\ {{\Delta}^{\perp 2} \over (1-\zeta)P^+}\right)\ ,
\end{eqnarray}
The four-momentum transfer from the target is
\begin{eqnarray}
\label{delta}   
\Delta&=&P-P'\ =\
\left( \zeta P^+\ ,\ {\Delta^\perp}\ ,\
{t+{\Delta^\perp}^2 \over \zeta P^+}\right)\ ,
\end{eqnarray}
where $t = \Delta^2$. In addition, overall energy-momentum
conservation requires $\Delta^- = P^- - P'^-$, which connects 
${\Delta^\perp}^2$, $\zeta$, and $t$ according to
\begin{equation}
 (1-\zeta) t  = -{\Delta^\perp}^2 .
 \label{tzeta}
\end{equation}  

The  details of the method of calculating $F^q$ and $\tilde F^q$ using the
Fock space expansion of the state was given in \cite{us1} in the limit of zero
$\zeta$. Here we do the calculation when both $\zeta$ and $\Delta^\perp$
are non-zero. There are quark and antiquark contributions
 respectively for $\zeta < x < 1$ and $-1 <x <\zeta-1$. In this work,
 we restrict ourselves to these two regions. 
The two-particle LFWFs for the photon can be calculated in perturbation
theory and are given by \cite{kundu}
\be
\psi_{2 s_1, s_2}^\lambda(x,q^\perp) &=& {1\over m^2-{m^2+{(q^\perp)}^2
\over x (1-x)}} {e e_q\over \sqrt{ 2 {(2 \pi)}^3}} 
\chi^\dagger_{s_1} \Big [
{(\sigma^\perp \cdot q^\perp)\over x} \sigma^\perp \nonumber\\&&- \sigma^\perp
{(\sigma^\perp \cdot q^\perp)\over 1-x} -i {m \over x (1-x)} \sigma^\perp
\Big ] \chi_{-s_2} \epsilon^{\perp *}_{\lambda} 
\ee
where we have used the two-component formalism \cite{two,kundu} and $m$ 
is the mass of $q(\bar{q})$. $\lambda$ is the helicity of the photon and 
$s_1,s_2$ are the helicities of the $q$ and ${\bar q}$ respectively. The
wave functions are expressed in terms of Jacobi momenta $x_i={k_i^+\over P^+}$ and
$q_i^\perp=k_i^\perp-x_i P^\perp$. These obey the relations $\sum_i x_i=1,
\sum_i q_i^\perp=0$.  
The GPDs can be written in terms of the overlaps of the LFWFs as follows :
\be 
F^q &=& \int d^2 q^\perp dx_1 \delta(x-x_1) \psi_2^*(\tilde
x,q^\perp-(1-\tilde x) 
\Delta^\perp
)\psi_2(x_1, q^\perp)\nonumber\\&&~~-\int d^2 q^\perp dx_1 
\delta(1+x-x_1) \psi_2^*(\tilde x,q^\perp+\tilde x \Delta^\perp
)\psi_2(x_1, q^\perp) 
\ee

\begin{figure}
\centering
\mbox{\subfigure{\tiny{(a)}\includegraphics[width=8cm,height=6cm,clip]{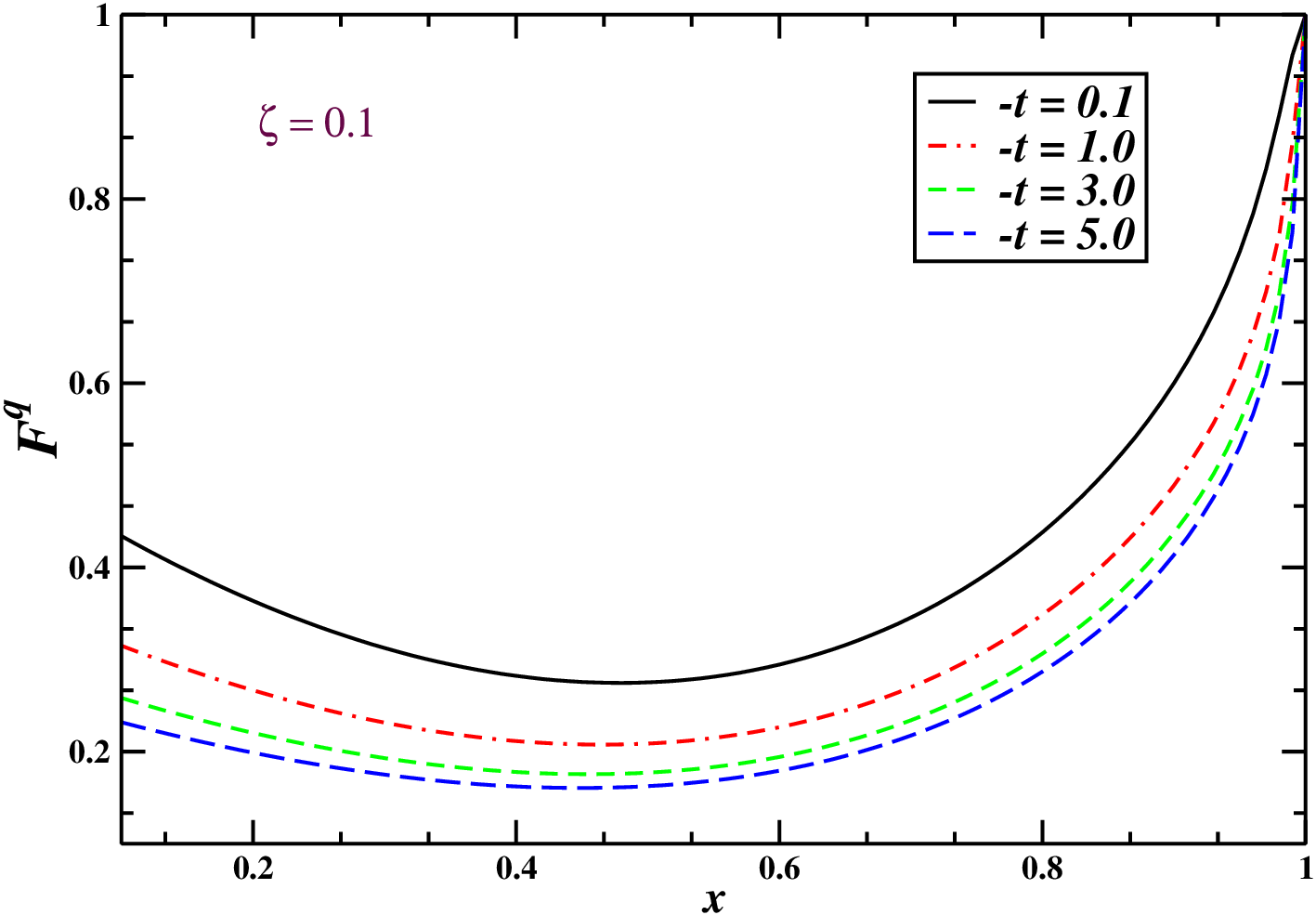}
\quad
\subfigure{\tiny{(b)}\includegraphics[width=8cm,height=6cm,clip]{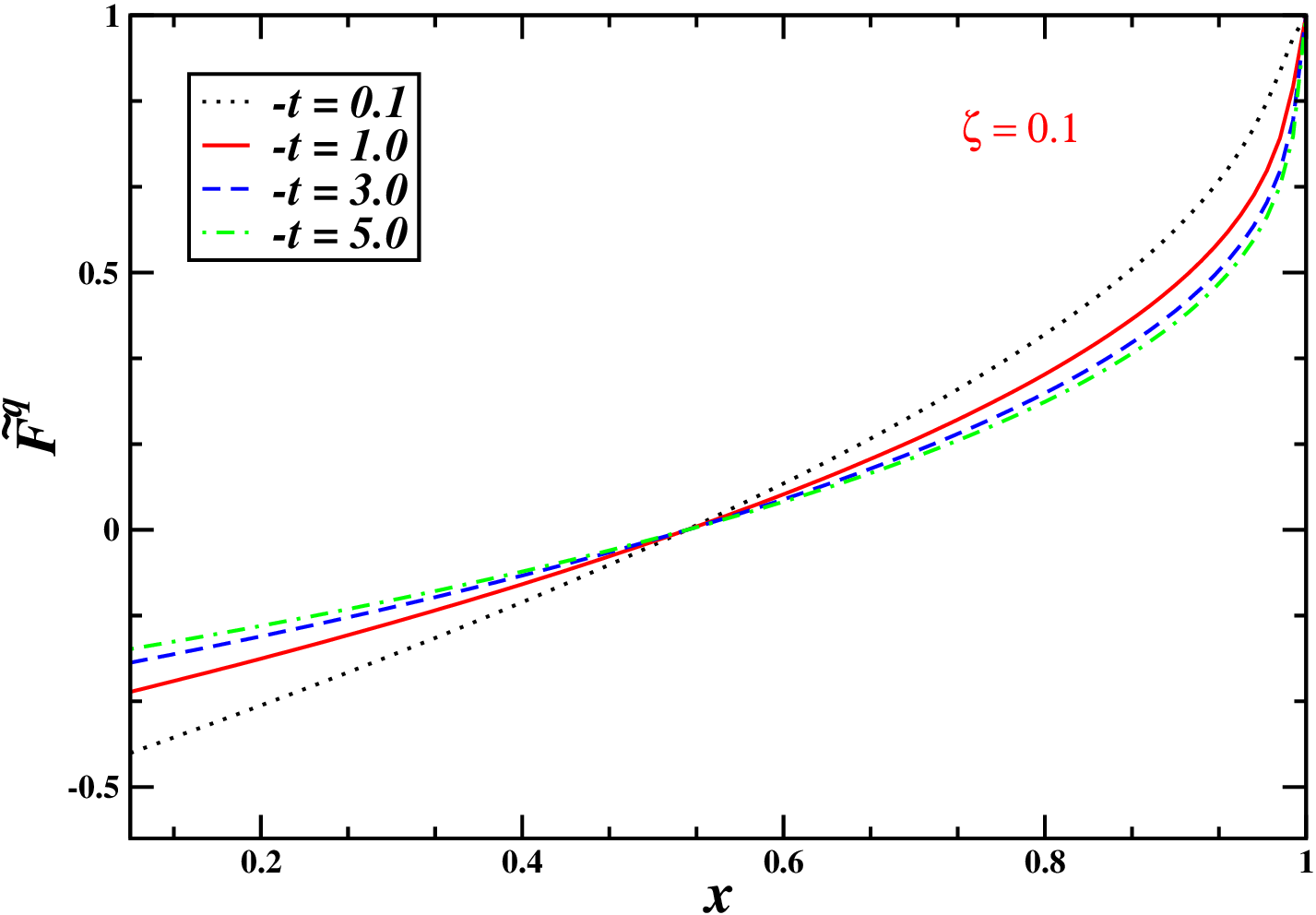} }}}

\centering
\mbox{\subfigure{\tiny{(c)}\includegraphics[width=8cm,height=6cm,clip]{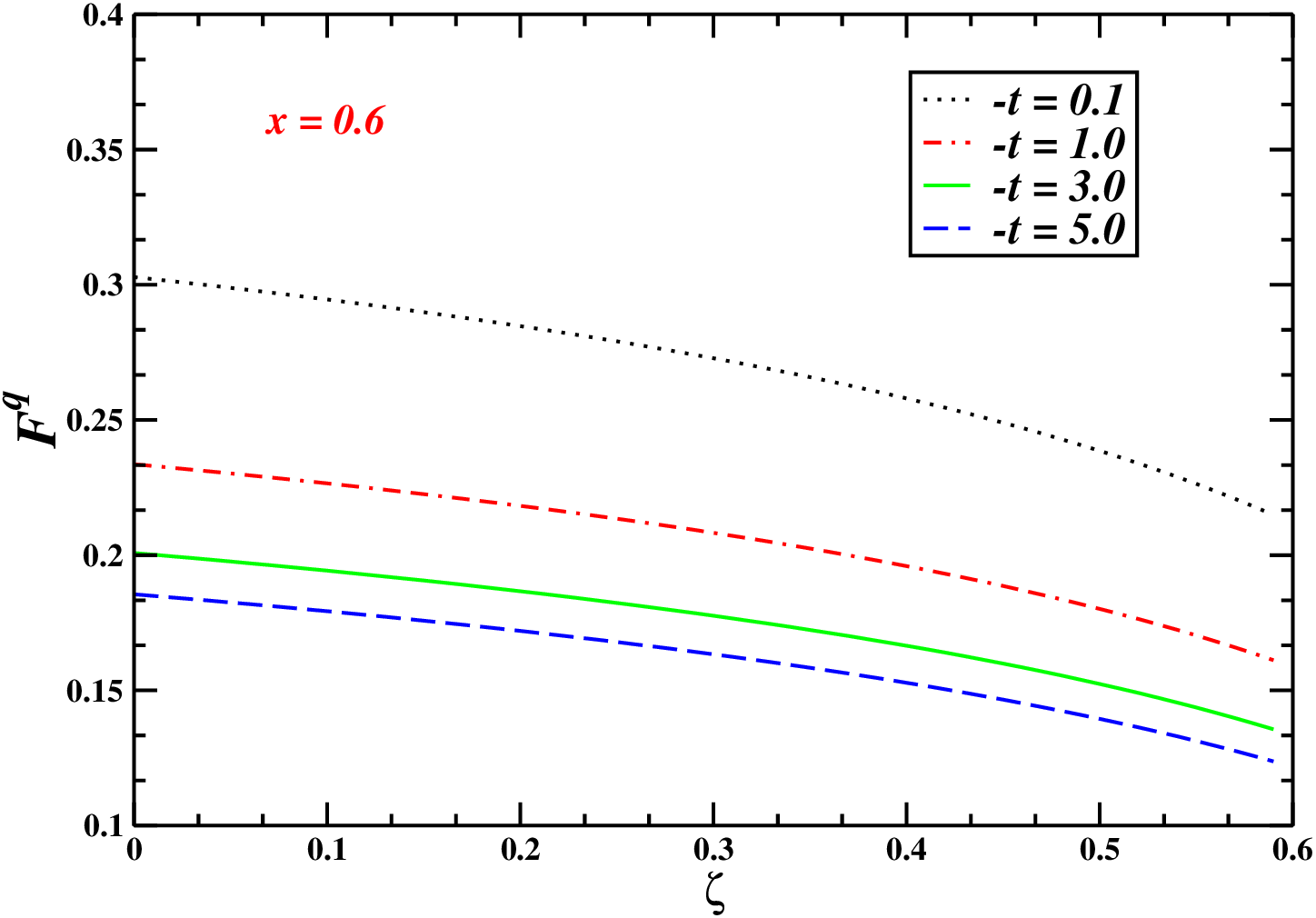}
\quad
\subfigure{\tiny{(d)}\includegraphics[width=8cm,height=6cm,clip]{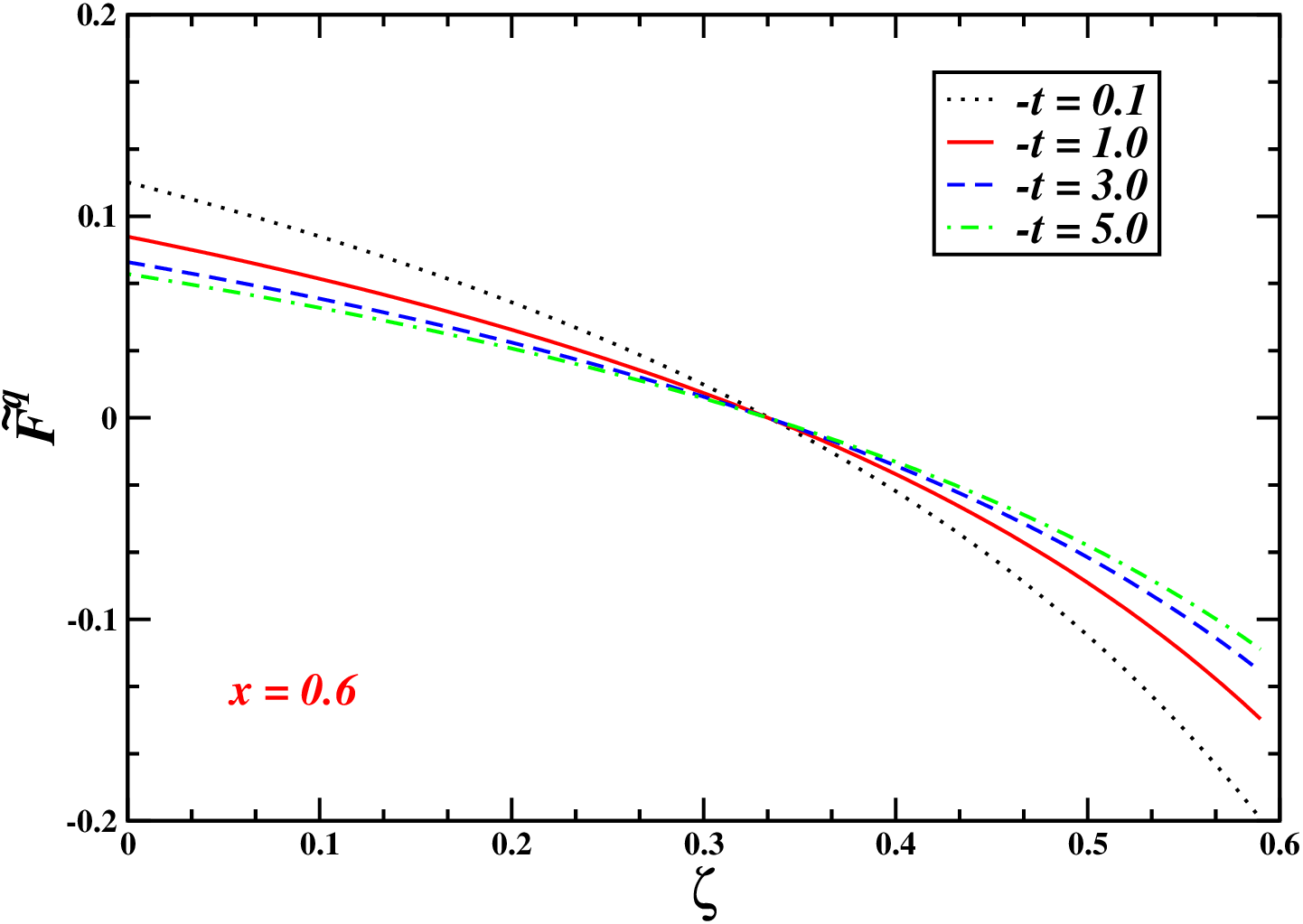} }}}

\caption{(Color online)  (a) Plot of unpolarized
GPD $F^{q}$ vs $x$ for fixed values of $-t$ in $GeV^{2}$ and at $\zeta=0.1$
and (b) polarized GPD  $\tilde{F^{q}}$ vs $x$ for fixed values of $-t$ in
$GeV^{2}$ and at $\zeta=0.1$ and (c) unpolarized
GPD $F^{q}$ vs $\zeta$ for fixed values of $-t$ in $GeV^{2}$ and at $x=0.6$
and (d) polarized GPD  $\tilde{F^{q}}$ vs $\zeta$ for fixed values of $-t$ in
$GeV^{2}$ and at $x=0.6$ , $\Lambda = 20 \mathrm{GeV}$. }
\label{fig1}

\end{figure}


Here $\tilde x={x_1-\zeta\over 1-\zeta}$. We have suppressed the helicity indices
and the sum over them. The first
term is the contribution from the quarks and the second is the
contribution from the antiquark in the photon. As the light-cone momentum
fraction $x_1$ has to be always greater than zero, the first term
contributes when $1>x>\zeta$ and the second term for $-1<x<\zeta-1$. 
We have taken $\zeta$ to be positive. There is a particle number changing
overlap in the region $0<x<\zeta$, Using the LFWFs each component can be 
calculated separately. We calculate in the same
reference frame as \cite{hadron_optics}. Note that the light cone plus
momentum of the target photon is non-zero. Finally we get for the unpolarized
photon


\begin{figure}
\centering
\mbox{\subfigure{\tiny{(a)}\includegraphics[width=8cm,height=6cm,clip]{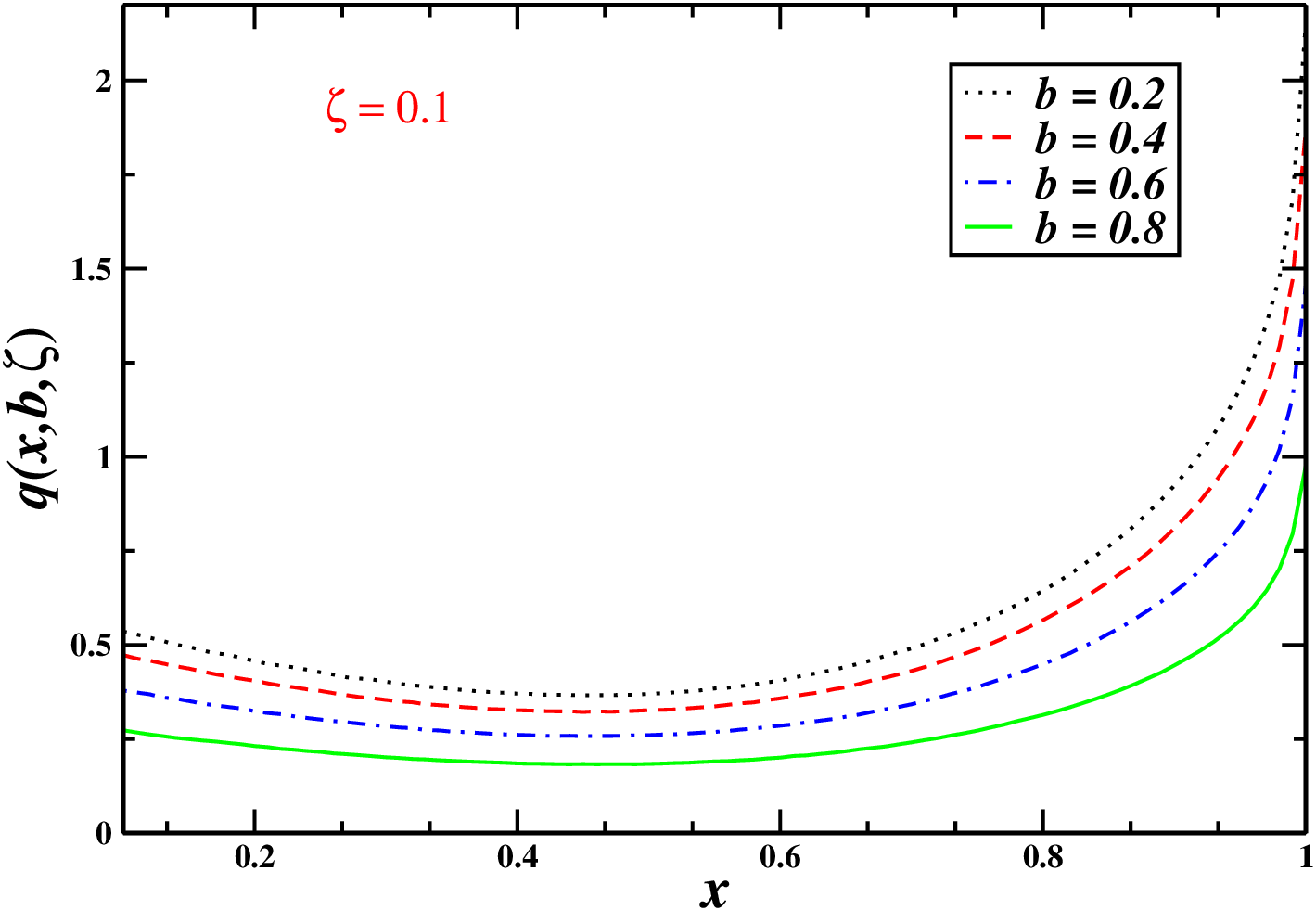}
\quad
\subfigure{\tiny{(b)}\includegraphics[width=8cm,height=6cm,clip]{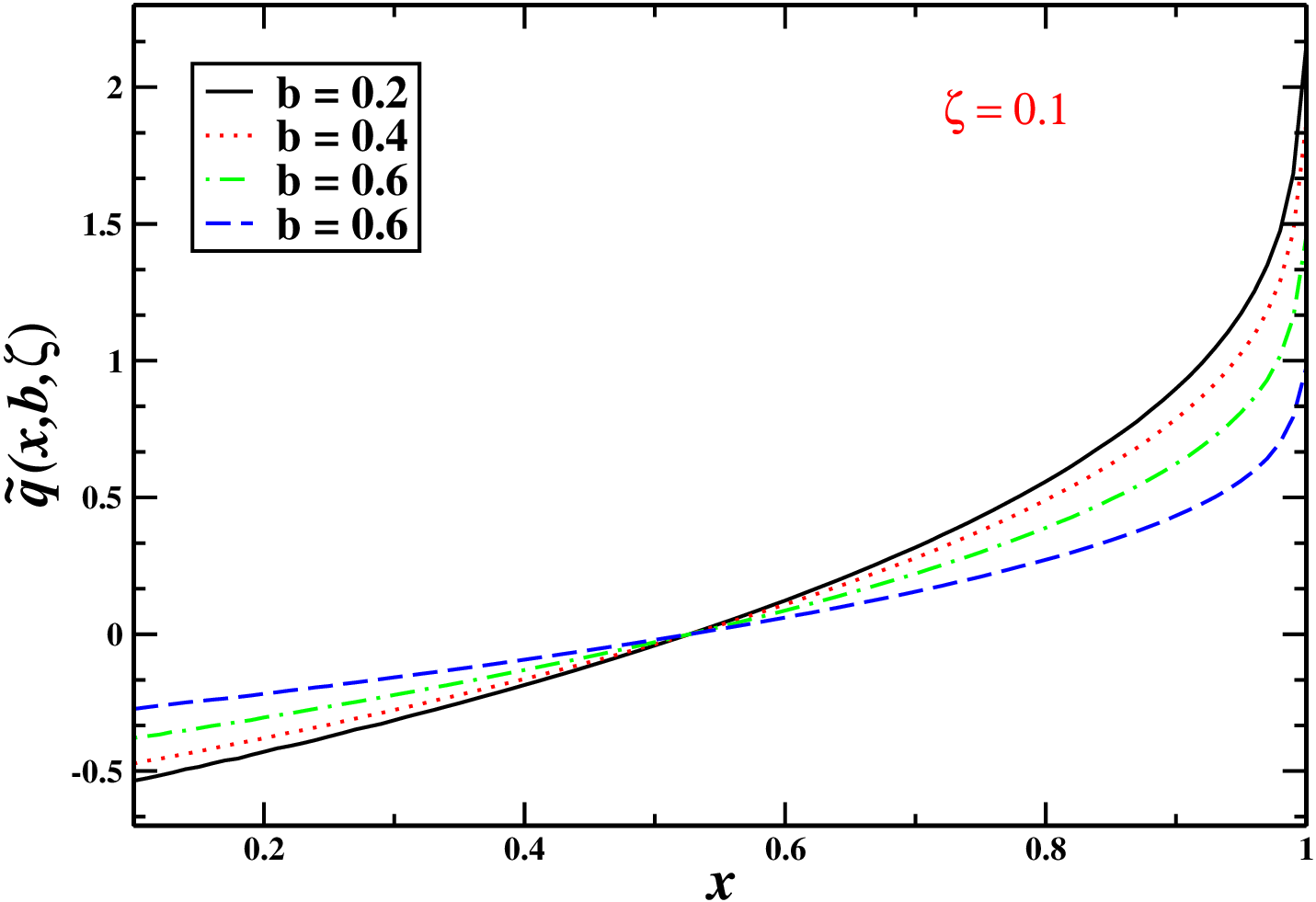} }}}

\centering
\mbox{\subfigure{\tiny{(c)}\includegraphics[width=8cm,height=6cm,clip]{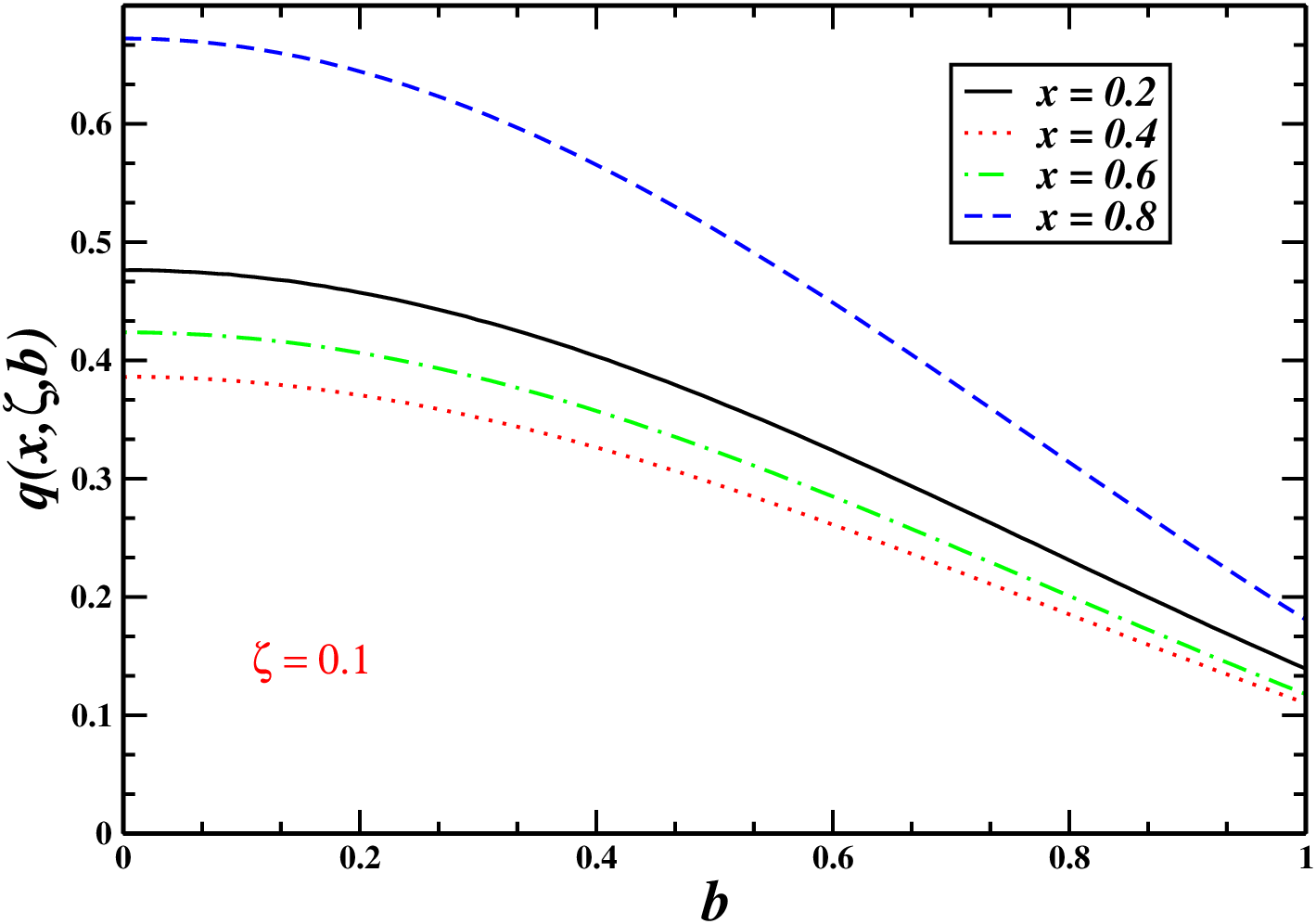}
\quad
\subfigure{\tiny{(d)}\includegraphics[width=8cm,height=6cm,clip]{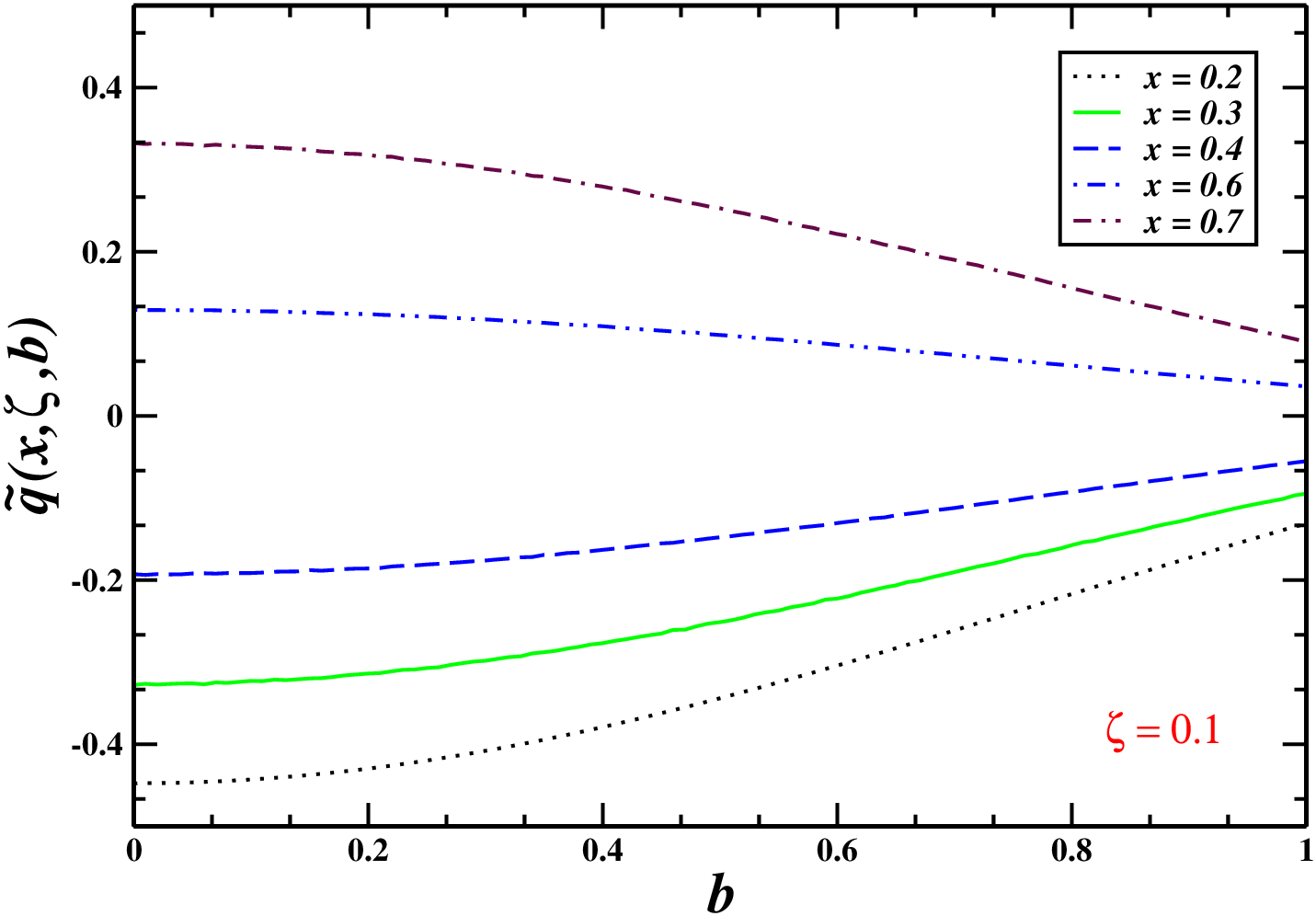} }}}
\caption{(Color online) (a) Plot of $q(x,\zeta,b)$ vs $x$  and (b) $\tilde
q(x,\zeta,b)$ vs $x$ for fixed values of $b$ and at $\zeta = 0.1$ 
(c) Plot of $q(x,\zeta,b)$ vs $b$  and (d) $\tilde q(x,\zeta,b)$ vs $b$ for fixed
 values of $x$ and at $\zeta = 0.1$,where we have taken 
$\Lambda$ = 20 $\mathrm{GeV}$ and $\Delta_{max}$= 3GeV where $\Delta_{max}$ is the upper limit in
the $\Delta$ integration. The distributions are given in $\mathrm{GeV}^2$ and 
$b$ is in ${\mathrm{GeV}}^{-1}$. }
\label{fig2}

\end{figure}


\be
F^q &=& \sum_q {\alpha e_q^2 \over 4 {\pi}^2 } \Big [ (1-x'-x+2 x' x) (I_1+I_2+L
I_3) +2 m^2 I_3 \big ] \theta(1-x) \theta(x-\zeta)  
\nonumber\\&&~~-\sum_q {\alpha e_q^2 \over 4 {\pi}^2 } 
\Big [ (-x-x"+2 (1+x) x") (I'_1+I'_2+ L'
I'_3) +2 m^2 I'_3 \Big ] \theta(1+x) \theta(1+x-\zeta);
\ee
where $x'={x-\zeta \over 1-\zeta}$ and $x"={1+x-\zeta \over 1-\zeta}$. 
Here the sum indicates sum over different quark flavors; $L=-2 m^2 + m^2 x
(1-x)+m^2 x' (1-x')- {(\Delta^\perp)}^2 (1-x')^2$, 
$L'=-2 m^2 - m^2 x (1+x)+ m^2 x"(1-x")  - {(\Delta^\perp)}^2 x"^2$; the 
integrals can be written as,


\begin{figure}
\centering
\mbox{\subfigure{\tiny{(a)}\includegraphics[width=8cm,height=6cm,clip]{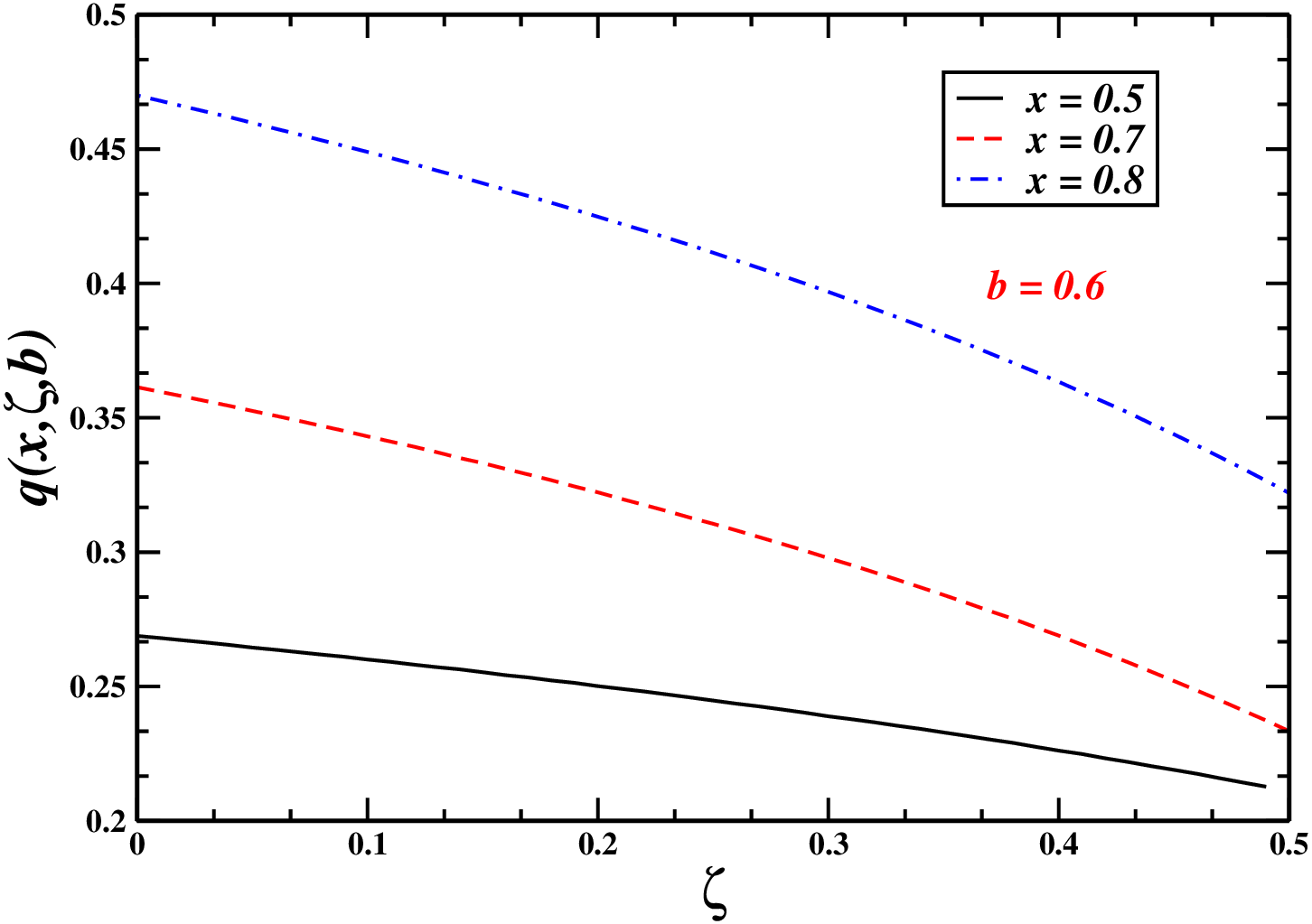}
\quad
\subfigure{\tiny{(b)}\includegraphics[width=8cm,height=6cm,clip]{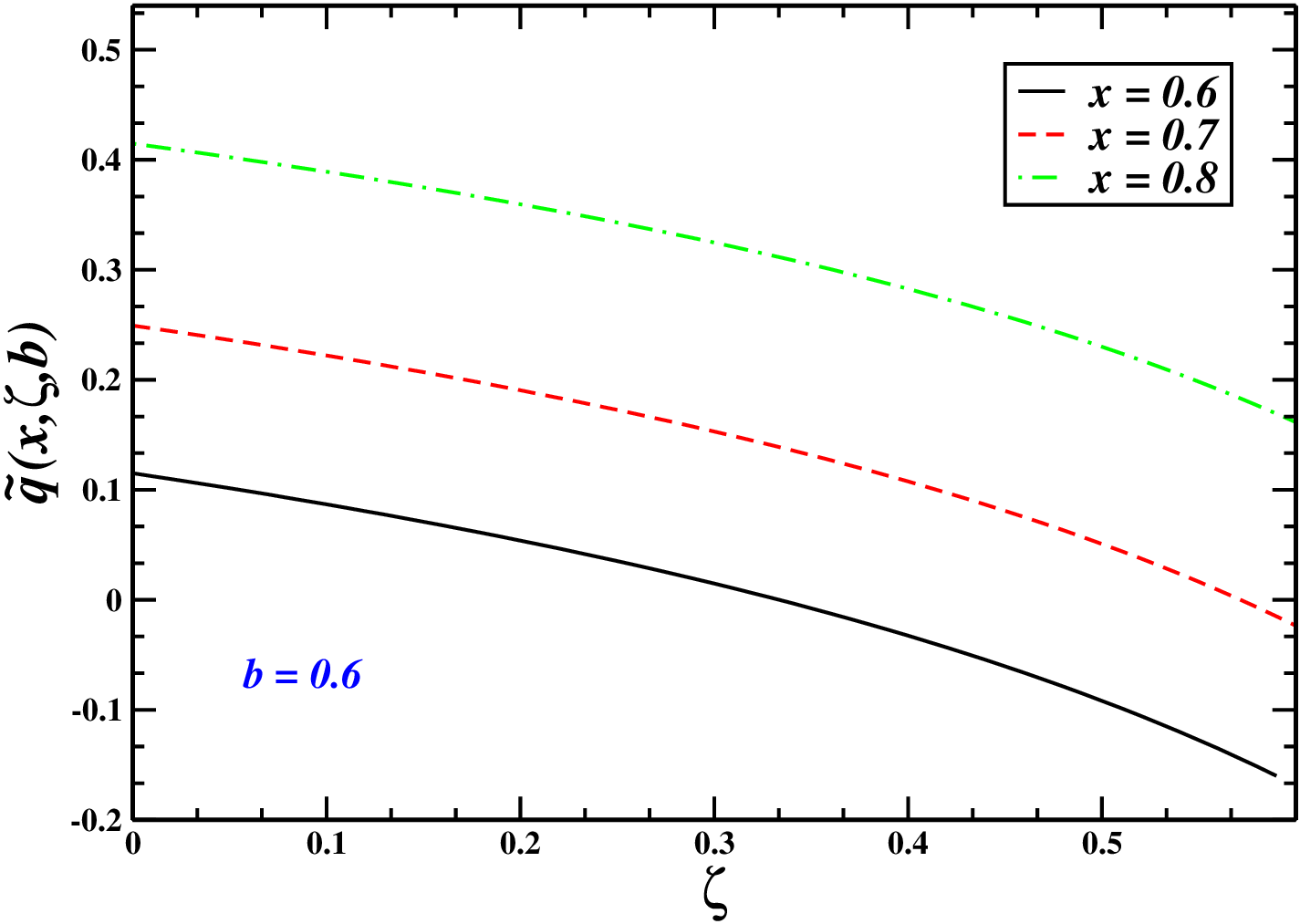} }}}

\centering
\mbox{\subfigure{\tiny{(c)}\includegraphics[width=8cm,height=6cm,clip]{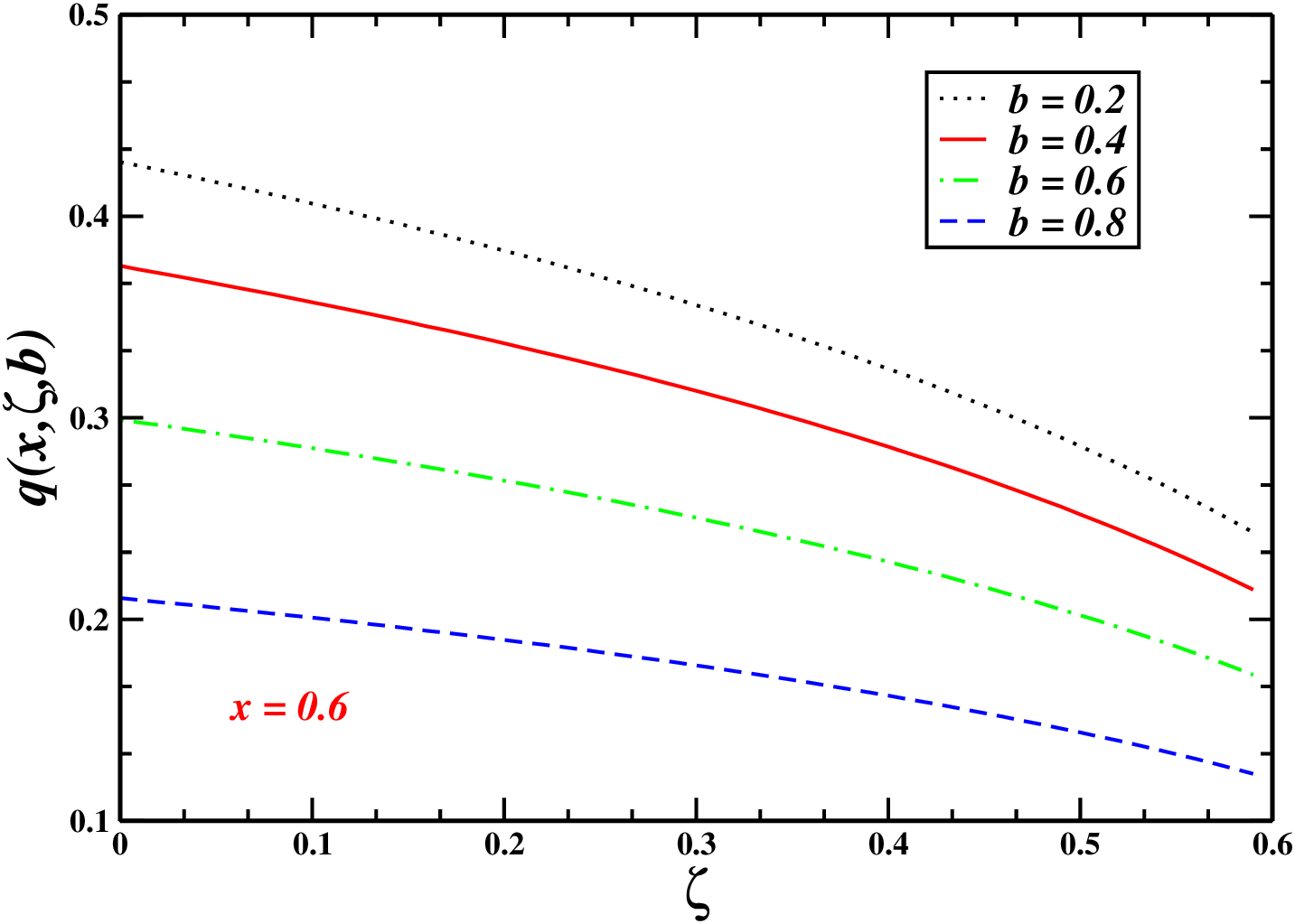}
\quad
\subfigure{\tiny{(d)}\includegraphics[width=8cm,height=6cm,clip]{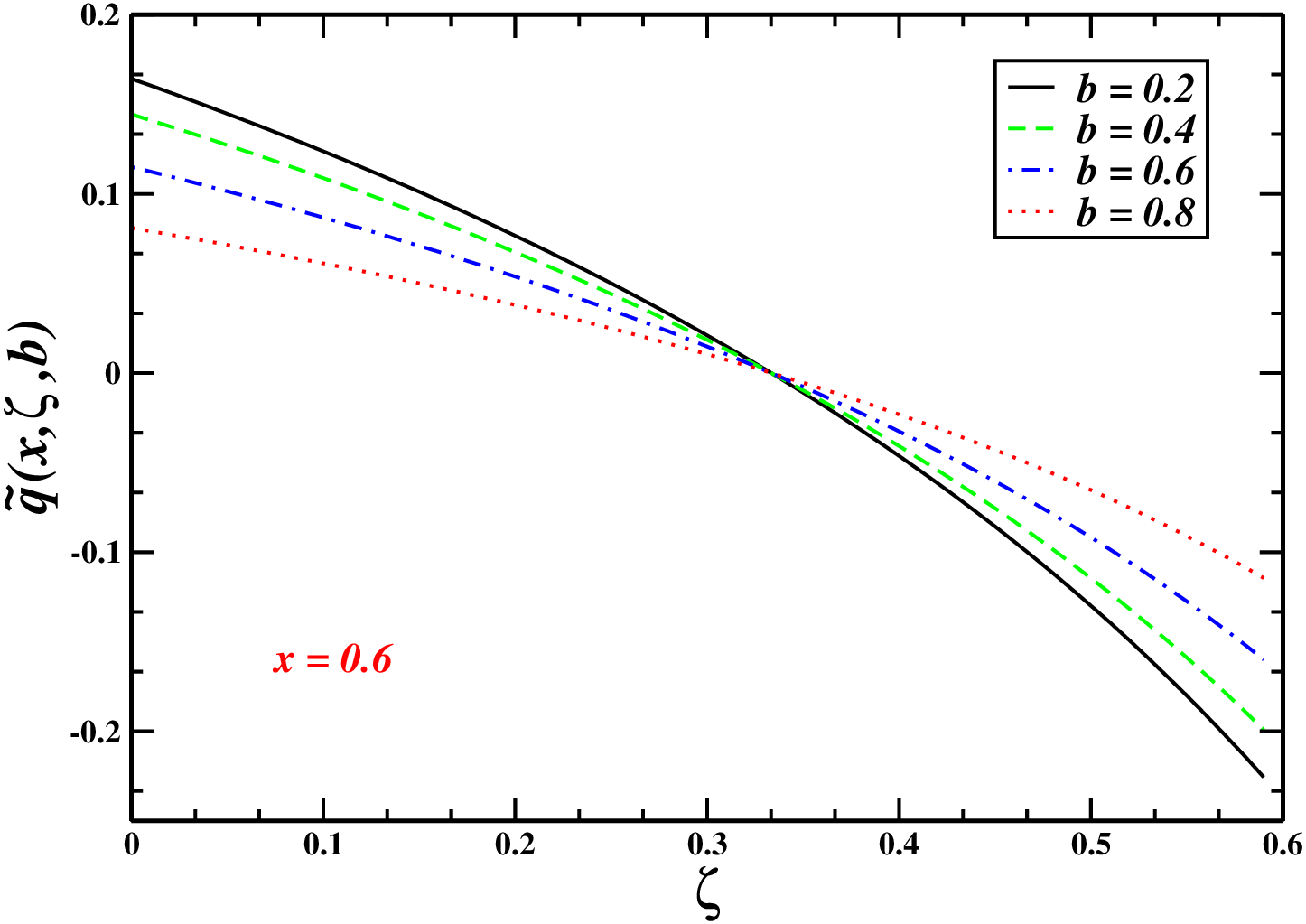} }}}
\caption{(Color online) (a) Plot of $q(x,\zeta,b)$ vs $\zeta$  and (b)
 $\tilde q(x,\zeta,b)$ vs $\zeta$ for fixed values of $x$ and at $b=0.6$ 
(c) Plot of $q(x,\zeta,b)$ vs $\zeta$  and (d) $\tilde q(x,\zeta,b)$ vs $\zeta$ for fixed values of $b$ and at $x=0.6$,where we have taken 
$\Lambda$ = 20 $\mathrm{GeV}$ and $\Delta_{max}$= 3GeV where $\Delta_{max}$ is the upper limit in
the $\Delta$ integration. The distributions are given in $\mathrm{GeV}^2$ and 
$b$ is in ${\mathrm{GeV}}^{-1}$.}
\label{fig3}

\end{figure}

\be
I_1=\int {d^2 q^\perp \over D} = \pi Log \Big [{\Lambda^2\over \mu^2-m^2 x (1-x)
+m^2}\Big ]\nonumber\\
I_2=\int {d^2 q^\perp \over D'} = \pi Log \Big [{\Lambda^2\over \mu^2-m^2 x'
(1-x')+m^2}\Big ]\nonumber\\
I_3= \int {d^2 q^\perp\over D D'}= \int_0^1 d \alpha {\pi\over P(x, \alpha,
 \zeta, {(\Delta^\perp)}^2)}
\ee 
where $D={(q^\perp)}^2 -m^2 x (1-x) +m^2$ and $D'= {(q^\perp)}^2
+{(\Delta^\perp)}^2 (1-x')^2 -2 q^\perp \cdot \Delta^\perp (1-x') -m^2 x'
(1-x')+m^2$, and $P(x, \alpha, \zeta, {(\Delta^\perp)}^2)= -\alpha m^2 x
(1-x)-m^2 (1-\alpha) x' (1-x') +m^2 + \alpha(1-\alpha) (1-x')^2 
{(\Delta^\perp)}^2$. As already seen in our previous paper for zero $\zeta$,
at zeroth order in $\alpha_s$ the
results are scale dependent, this scale dependence in our approach comes
from the upper limit of the transverse momentum integration $ \Lambda = Q$.
$\mu$ is a lower cutoff on the transverse momentum, which can be taken to
zero as long as the quark mass is nonzero. 
We included the subdominant contributions from the mass terms in the vertex.


\begin{figure}
\centering
\mbox{\subfigure{\tiny{(a)}\includegraphics[width=8cm,height=6cm,clip]{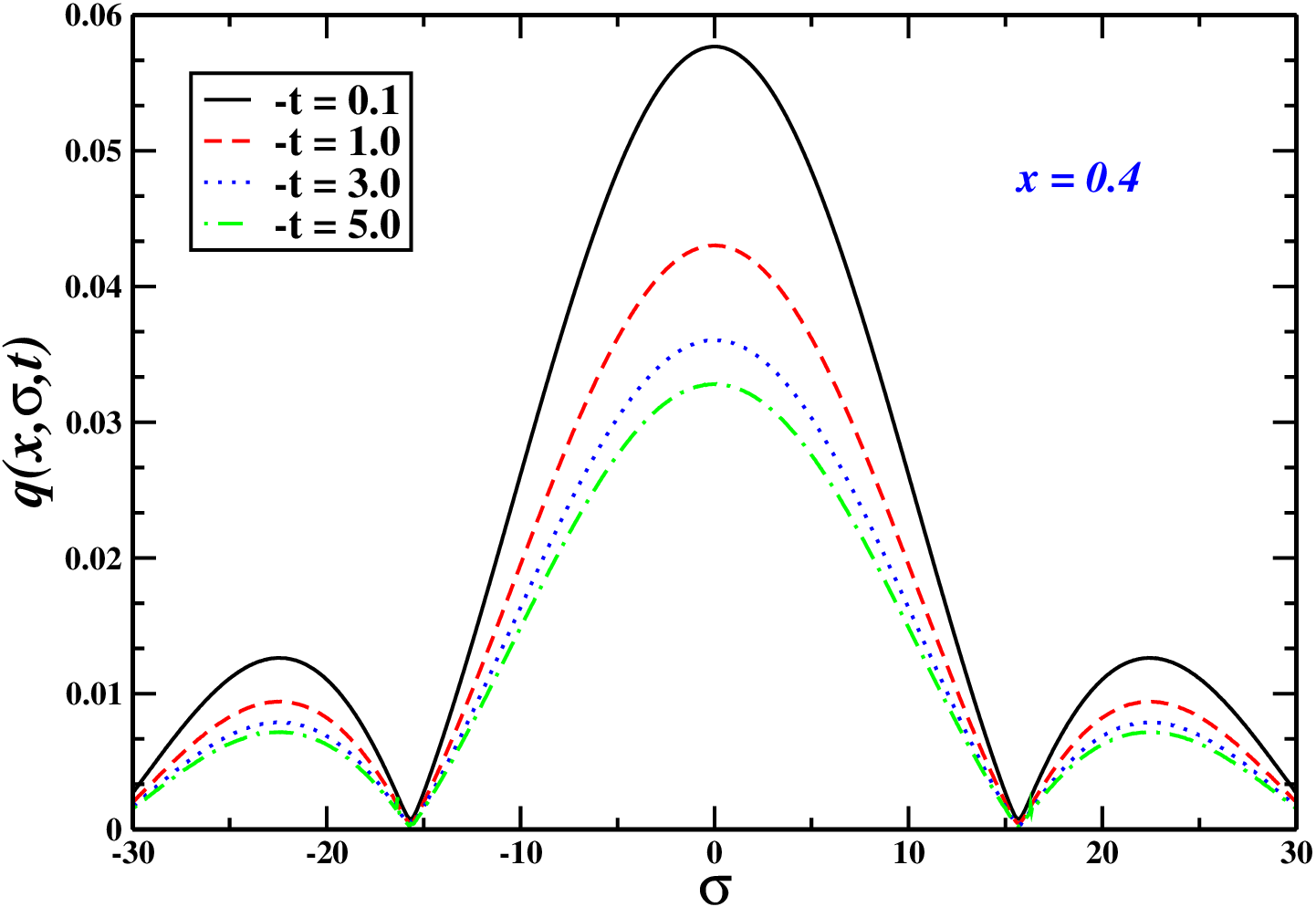}
\quad
\subfigure{\tiny{(b)}\includegraphics[width=8cm,height=6cm,clip]{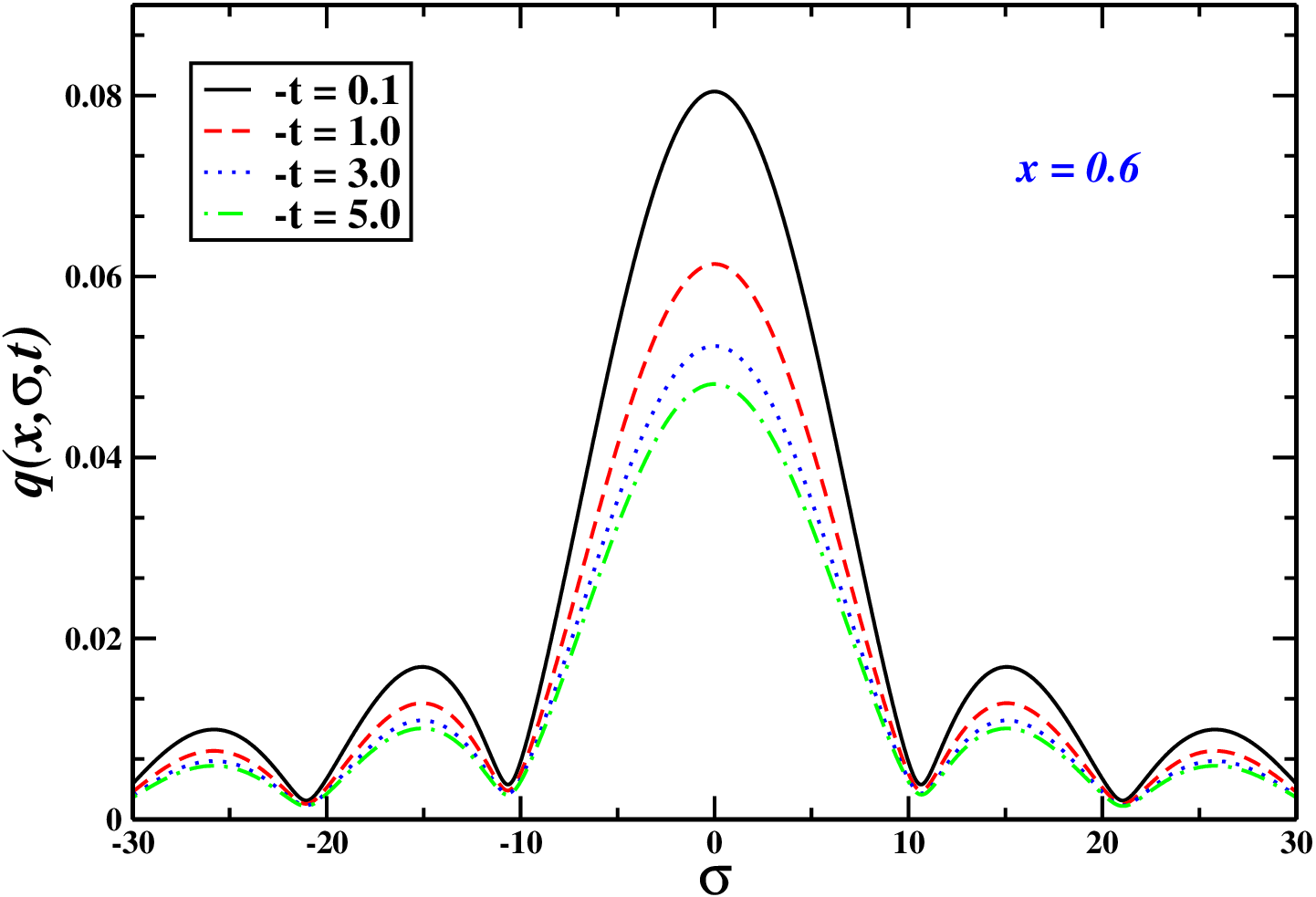} }}}

\centering
\mbox{\subfigure{\tiny{(c)}\includegraphics[width=8cm,height=6cm,clip]{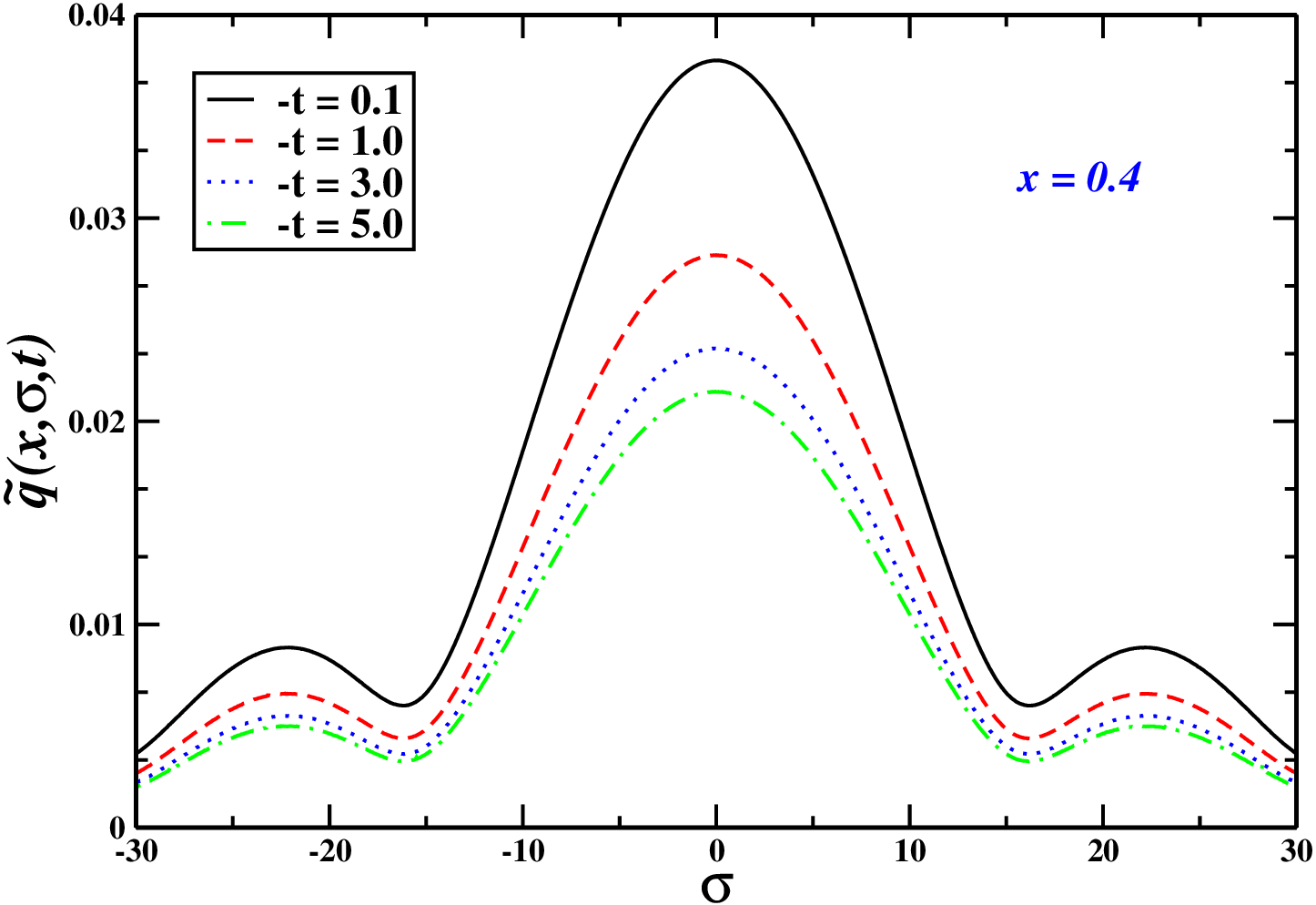}
\quad
\subfigure{\tiny{(d)}\includegraphics[width=8cm,height=6cm,clip]{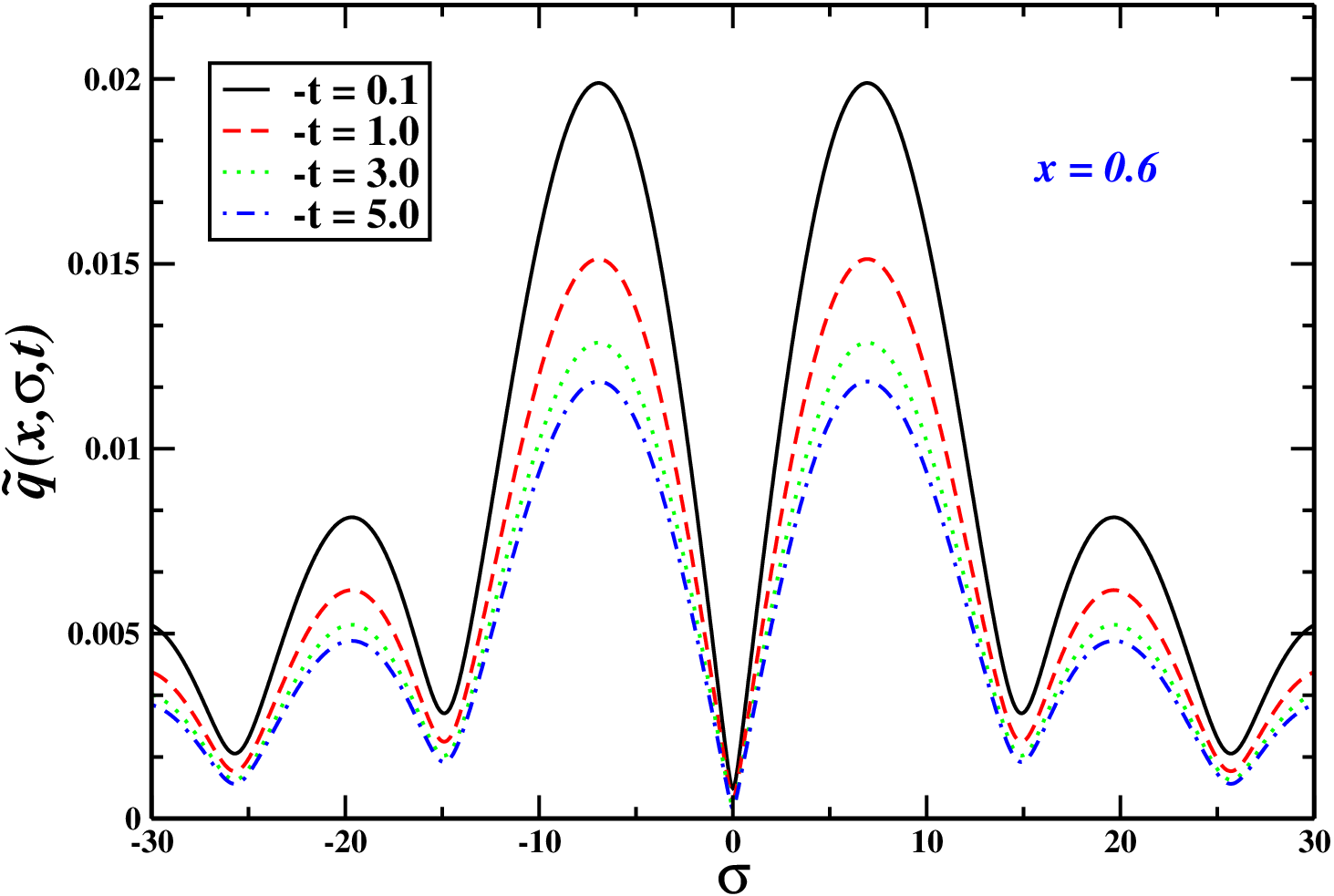} }}}
\caption{\label{fig2}(Color online) (a) Plot of $q(x,\sigma,t)$ vs $\sigma$  for
 a fixed value of $x=0.4$ and different values of $-t$ in $GeV^{2}$, (b) same
as (a) but for $x=0.6$; (c) $\tilde q(x,\sigma,t )$ vs $\sigma$ for a fixed
value of $x=0.4$ and different values of $-t$ in $GeV^{2}$, (d) same as (c) but
for $x=0.6$; we have taken $\Lambda$ = 20 $\mathrm{GeV}$.}

\end{figure}


For the antiquark contributions we have similar integrals
\be
I'_1=\int {d^2 q^\perp \over H} = \pi Log \Big [{\Lambda^2\over \mu^2
+m^2 x(1+x)+m^2}\Big ]\nonumber\\
I'_2=\int {d^2 q^\perp \over H'} = \pi Log \Big [{\Lambda^2\over \mu^2
-m^2 x"(1-x")+m^2}\Big ]\nonumber\\
I'_3= \int {d^2 q^\perp\over H H'}= \int_0^1 d \alpha {\pi\over Q(x, \alpha,
\zeta, {(\Delta^\perp)}^2)}
\ee

where $H={(q^\perp)}^2 +m^2 x (1+x) +m^2$ and $H'= {(q^\perp)}^2
+{(\Delta^\perp)}^2 x"^2 +2 q^\perp \cdot \Delta^\perp x" -m^2 x" (1-x")
+m^2$, and $Q(x, \alpha, \zeta, {(\Delta^\perp)}^2)= \alpha m^2 x (1+x) -
(1-\alpha) x" (1-x")m^2 + m^2 
+ \alpha(1-\alpha) x"^2 {(\Delta^\perp)}^2$.

For polarized photon the GPD $\tilde F^q$ can be calculated from the 
terms of the form 
$\epsilon^2_\lambda \epsilon^{1*}_\lambda-\epsilon^1_\lambda
\epsilon^{2*}_\lambda$ \cite{pire}. We consider the terms where the photon
helicity is not flipped. This can be written as, 
\be
\tilde F^q &=& \sum_q {\alpha e_q^2 \over 4 {\pi}^2 } \Big [ (x+ x' -1) (I_1+I_2+L
I_3) +2 m^2 I_3 \big ] \theta(1-x) \theta(x-\zeta) 
\nonumber\\&&+\sum_q {\alpha e_q^2 \over 4 {\pi}^2 } \Big [ 
(-x-x") (I'_1+I'_2+ L'
I'_3) +2 m^2 I'_3 \Big ] \theta(1+x) \theta(1+x-\zeta)
\ee

In the limit $\Delta^\perp=0$ we get the expressions in \cite{pire} and the
well known splitting functions in the forward limit. Starting from the
expressions of photon GPDs, we define the 
parton distributions of the photon  in impact parameter space for nonzero
$\zeta$ as,
\be
q(x,\zeta,b)={1\over 4 \pi^2} \int d^2 \Delta^\perp e^{-i \Delta^\perp 
\cdot b^\perp} F^q (x,\zeta,t)\nonumber\\
\tilde q (x,\zeta,b)={1\over 4 \pi^2} \int d^2 \Delta^\perp  e^{-i \Delta^\perp
\cdot b^\perp} \tilde F^q (x,\zeta, t).
\ee

Here $b= \mid b^\perp \mid $ is the transverse  impact parameter which is a
measure of  the transverse distance between the struck quark and the
center of momentum of the photon. In  the limit of zero skewness $\zeta$ 
the parton distributions in impact parameter space have a probabilistic
interpretation \cite{burkardt}. However, in most DVCS experiments
 $\zeta$ is nonzero, and it is
of interest to investigate the GPDs in $b^\perp$ space with  
non-zero $\zeta$. As mentioned in the introduction, the physical
interpretation of the parton distributions in impact parameter space was
given for the proton in \cite{diehl1} when the
skewness $\zeta$ is nonzero. In this case  the
transverse location of the target proton is different before and after 
the scattering. This transverse shift depends on the
skewness $\zeta$ and $b$ and the information on the
transverse shift is not washed out even if the GPDs are integrated over
$x$ in the DVCS amplitude. On the other hand, the boost invariant 
 longitudinal impact parameter  $\sigma$ was first introduced
in \cite{hadron_optics} and it was shown that DVCS amplitude shows
a diffraction pattern in longitudinal impact parameter space.
GPDs for the proton when expressed in term of $\sigma$ 
also exhibit the similar diffraction pattern \cite{manohar2}. In the same way
 here we introduce the boost invariant
 longitudinal impact parameter conjugate to the longitudinal momentum
 transfer as $\sigma=\frac{1}{2}b^-P^+$.
The photon GPDs in longitudinal position space is given by:
\be
q(x,\sigma, t)=\frac{1}{2\pi}\int_0^{\zeta_{max}} d\zeta e^{i \zeta P^+b^-/2}
F^q (x,\zeta,t)
= \frac{1}{2\pi}\int_0^{\zeta_{max}} d\zeta e^{i \zeta \sigma} 
F^q (x,\zeta,t)\nonumber\\   
\tilde q (x,\sigma, t)=\frac{1}{2\pi}\int_0^{\zeta_{max}}
 d\zeta e^{i\zeta P^+b^-/2}\tilde F^q (x,\zeta,t)
= \frac{1}{2\pi}\int_0^{\zeta_{\max}} d\zeta e^{i \zeta \sigma} \tilde F^q (x,\zeta,t)
\ee
In the numerical calculation, we shall consider  only in the region
 $\zeta<x<1$, and the upper limit of $\zeta$ integration $\zeta_{max}$ will be
taken as $x$.

\begin{figure}
\centering
\mbox{\subfigure{\tiny{(a)}\includegraphics[width=8cm,height=6cm,clip]{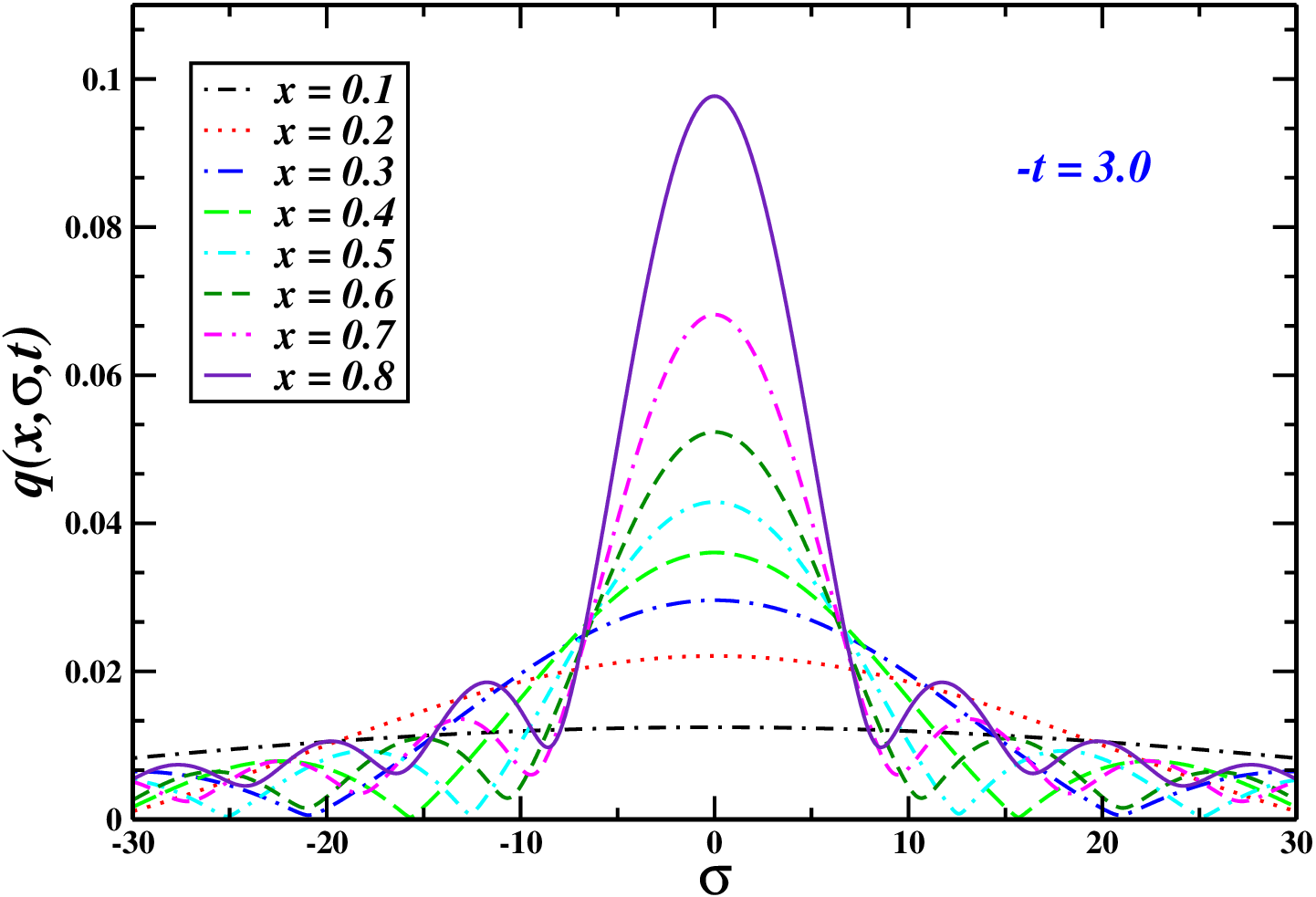}
\quad
\subfigure{\tiny{(b)}\includegraphics[width=8cm,height=6cm,clip]{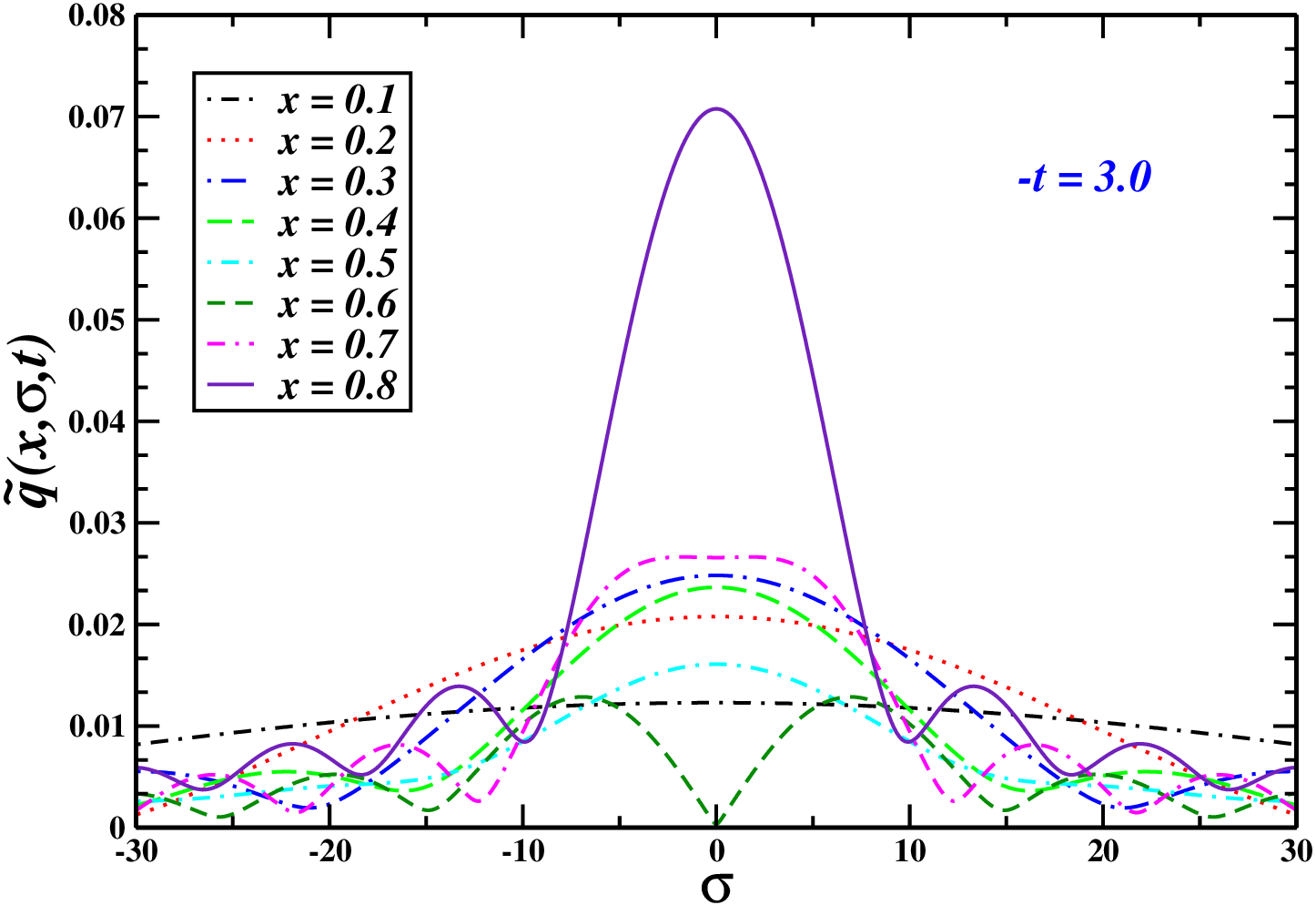} }}}

\centering
\mbox{\subfigure{\tiny{(c)}\includegraphics[width=8cm,height=6cm,clip]{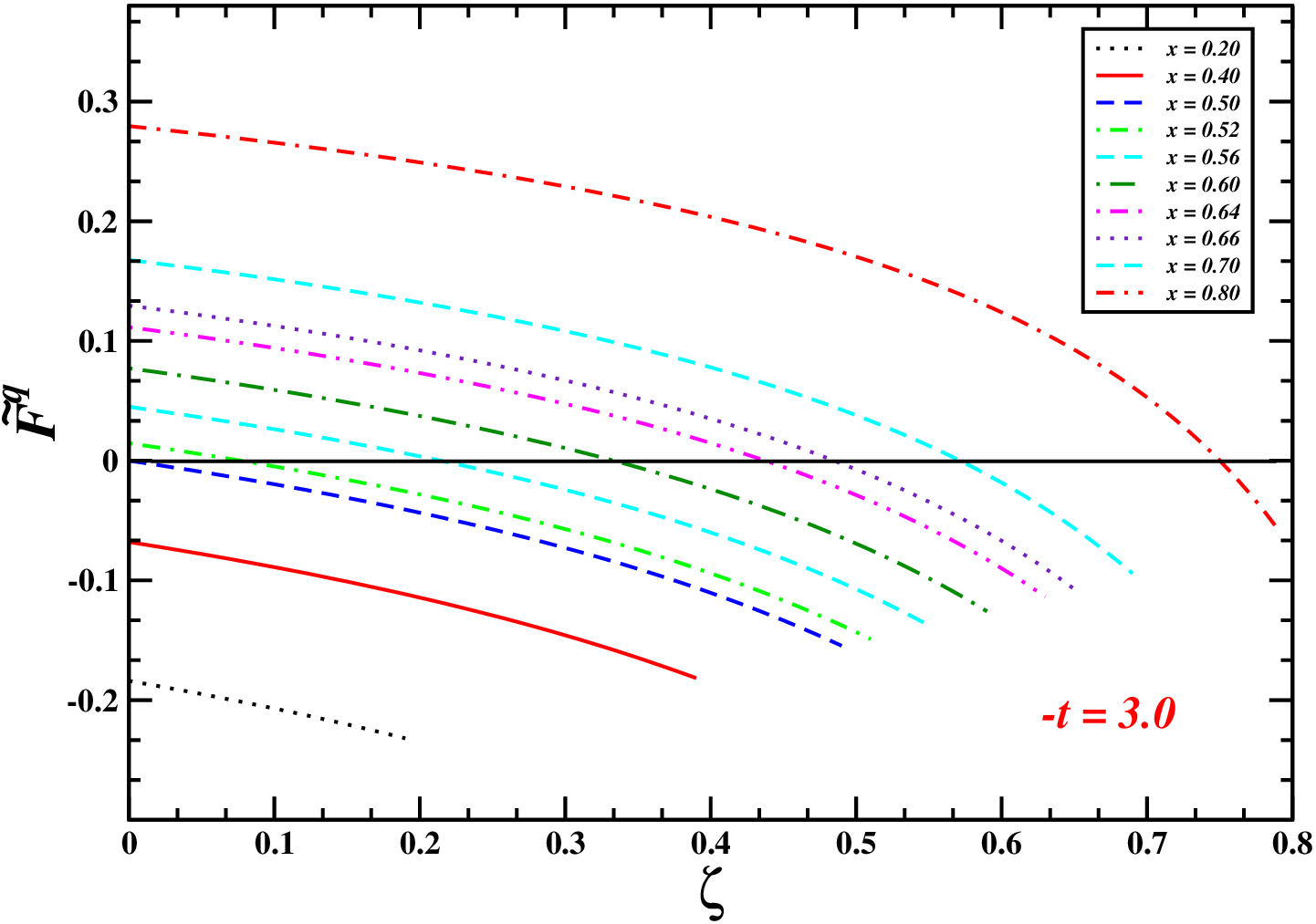}
\quad
\subfigure{\tiny{(d)}\includegraphics[width=8cm,height=6cm,clip]{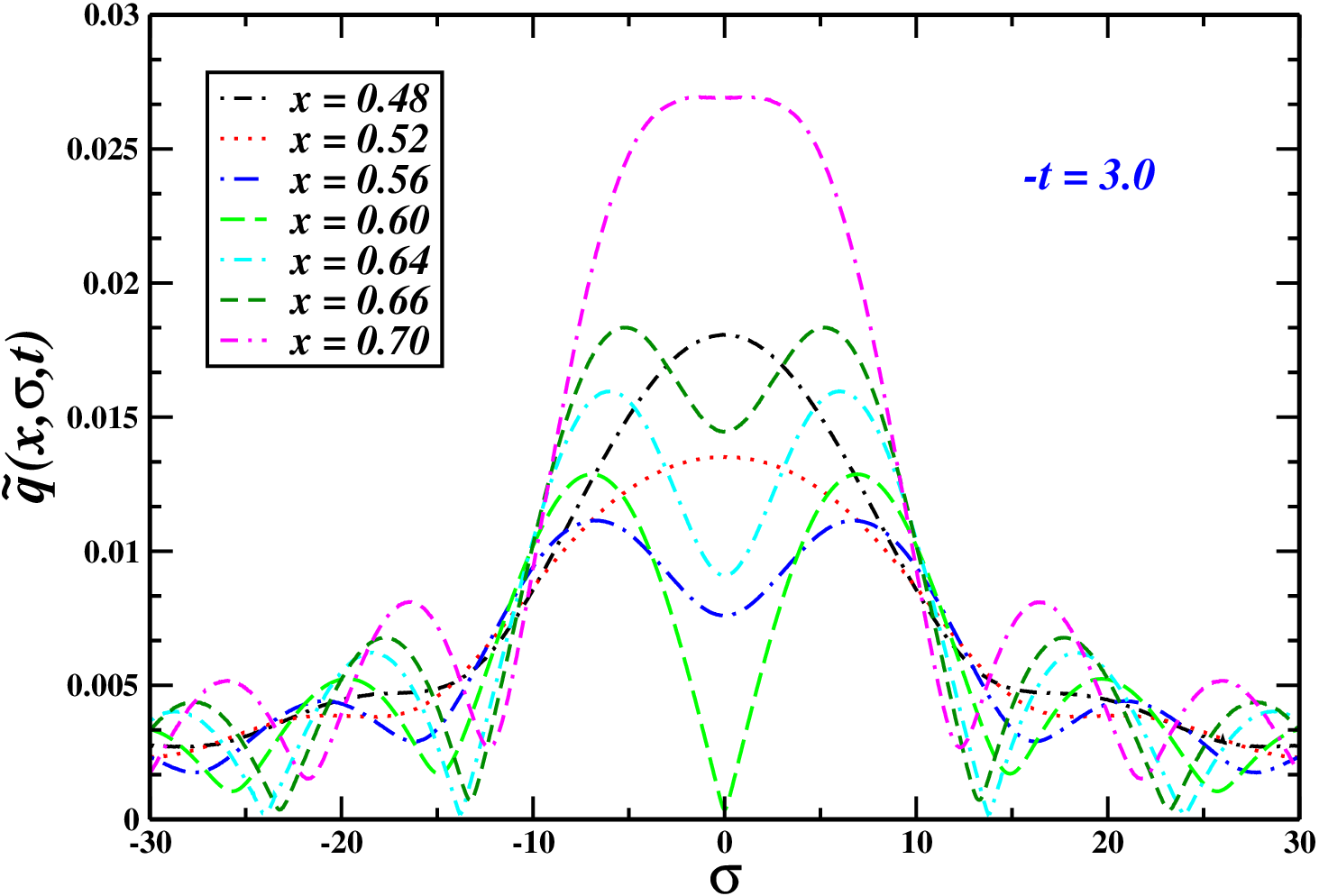} }}}

\caption{\label{fig5}(Color online) Plots of (a) $q(x,\sigma,t)$ and (b)
$\tilde{q} (x, \sigma, t)$  vs $\sigma$ for fixed $t$ in ${GeV}^2$ and
different values of $x$ ; (c) $\tilde{F^q} $ vs $\zeta$ for $-t= 3.0 $ ${GeV}^2$  
and different values of $x$ ,(d) $\tilde{q} (x, \sigma, t)$ 
 vs $\sigma$ for fixed $t$ in ${GeV}^2$ and different values of
 $x$ for which $\tilde{q} (x, \sigma, t)$ shows two maxima with a central minima} 

\end{figure}


\section{Numerical Results}
In all plots we have taken $\Lambda$ =Q= 20 GeV and $m$=0.0033 GeV.  
In Figs. 1(a) and (b) we have plotted the GPDs for unpolarized photon $F^q$
and for polarized photon $\tilde F^q$ as functions of $x$ for a fixed value
of $\zeta$ and different values of $t$. Note that in all plots $x$ is
positive and $x>\zeta$ so that the contribution comes from the active quark 
in the photon. 
As seen in our previous work and also in \cite{pire} when $\Delta^\perp$ is zero, that is
when the momentum transfer is purely in the longitudinal direction, $F^q$ is
symmetric with respect to $x=1/2$ when the momentum is shared equally among
the quark and the antiquark. This is the case when the subleading terms due
to the quark mass at
the vertex  are neglected. When $\Delta^\perp$ is non-zero, this symmetry is
not there, the effect is more prominent at lower values of $x$. We see a
similar effect  when both $\zeta$ and $\Delta^\perp$ are non-zero. Both the
GPDs $F^q$ and $\tilde F^q$ have been normalized as in \cite{us1} to compare with
\cite{pire} in the limit of zero $t$. They become independent of $t$ as $x
\rightarrow 1$ because in this limit all the momentum is carried by the
quark in the photon. The slope of
 the polarized GPD changes with increase of $-t$ and at
$x \approx 0.5$ it changes sign. Figs 1(c) and 1(d) show the dependence of
$F^q$ and $\tilde F^q$ on $\zeta$ for a fixed value of $x$ and different
values of $t$. Both $F^q$ and $\tilde F^q$ decrease steadily with increase
of $\zeta$. The rate of decrease is faster for lower values of $\mid -t \mid
$. In Fig. 2 we have plotted the Fourier transform (FT) of $F^q$ and $\tilde
F^q$ with respect to $\Delta^\perp$ for non-zero $\zeta$. we took the upper
limit of the $\Delta^\perp$ integration, $\Delta_{max}$=3 GeV. Figs. 
2(a)- 2(d) show
the plots of the impact parameter dependent parton distributions
$q(x,\zeta,b)$ and $\tilde q(x,\zeta,b)$ both as functions of $x$ for fixed
$b$ as well as functions of $b$ for fixed $x$; for a given value of 
 $\zeta$.
As stated before in our calculation we have restricted ourselves to the
kinematical region  $x>\zeta$. As observed for $\zeta=0$, both $q(x,
\zeta,b)$ and $\tilde q(x,\zeta,b)$ rise sharply as $x \rightarrow 1$ as the
GPDs become independent of $t$. In Figs. 3(a)-(d) we have shown the 
behaviour of $q(x,\zeta,b)$ and $\tilde q(x,\zeta,b)$ as functions of
$\zeta$ for fixed $x$ and and $b=\mid b^\perp \mid$ values. Both decrease
steadily with increase of $\zeta$. For a given $b$, the rate of decrease 
with $\zeta$ increases with the increase of $x$ and for a given $x$ the rate 
decreases with increase of $b$. At a given $\zeta$, the distributions are
broader in $b$ space for $x \approx 1/2$ when both $q$ and $\bar{q}$ carry
equal momenta. As stated in the introduction, the GPDs in transverse impact
parameter space for non-zero $\zeta$ probe partons inside the target photon
when the initial photon is displaced from the final photon in the transverse
impact parameter space. The $x$ and $\zeta$  dependence of the photon GPDs 
play an important role in the longitudinal position space distribution which
we show in Fig 4. Fig. 4 (a) and (b) show the plot of $q(x,\sigma,t)$ vs. $\sigma$
for different values of $t$ and two different values of $x$. 
4 (c) and (d)  show the plots of $\tilde q(x,\sigma,t)$ vs. $\sigma$.
$q(x, \sigma, t)$ shows prominent diffraction pattern with a
central maximum and several secondary maxima separated by well-defined minima
at a fixed value of $x=0.4$. The diffraction pattern is not that prominent
for $\tilde q(x, \sigma,t)$ at the same value of $x$. Also for $q(x, \sigma,
t)$ the diffraction pattern becomes less prominent at higher values of $x$
and disappears at $x \approx 0.8$. $\tilde q(x, \sigma, t)$ show two major
peaks separated by a central minimum at $x=0.6$. 

In order to further study the diffraction pattern in $\sigma$ space in Figs.
5(a) and (b) we plot $q(x,\sigma,t)$ and $\tilde q(x, \sigma,t)$ respectively 
vs $\sigma$ for a fixed value of $t$ and different values of $x$. As can
been seen from the plots, the positions of the first minima for both move
towards smaller values of $\sigma$ as $x$ increases. We compare this
observation with the diffraction pattern shown by the proton GPDs in
$\sigma$ space and by the DVCS amplitude for a dressed electron. 
There the positions of the first minima moved in towards smaller values of
$\sigma$ with the increase of $-t$. In contrast, in the case of photon GPDs
the positions of the minima are independent of $-t$. An analogy with the
diffraction pattern in optics was given in \cite{hadron_optics}. It was
conjectured that the finite range of $\zeta$ integration acts as a slit of
finite width necessary to produce the diffraction pattern. $\zeta_{max}$ for
a given $t$ is determined by DVCS kinematics and is proportional to $(-t)$
for the proton target of finite mass. This $\zeta_{max}$ is analogous to 
the slit width in optics experiments. There the width of the central maximum 
is inversely proportional to the slit width, and in analogy, in DVCS the
width is inversely proportional to $\zeta_{\max}$ and therefore to $-t$.
However the photon is massless and as seen from Eq. (\ref{tzeta}), for a
photon target $\zeta_{max}$ is $1$ in the limiting sense when
${(\Delta^\perp)}^2=0$ for a given $t$. In our analysis, $\zeta_{max}=x$. 
So the slit width in the optics analogy is proportional to $x$ and as
expected the minima move to smaller values of $\sigma $ as $x$ increases. 
As seen in Figs. 4(d), 5(b) and 5(d), $\tilde q (x, \sigma, t)$ has two
maxima separated by a central minimum for those values of $x$ for which   
$\tilde F^q$ changes sign in the $\zeta$ range covered as seen in Fig. 5 (c).
At $x=0.6$, $\tilde F^q$ changes sign almost in the middle of the 
$\zeta$ range covered : it is positive for half of the $\zeta$ range and 
negative for the other half as a result $\tilde q (x, \sigma, t)$ has the
lowest value for the minimum at $x=0.6$. Also as seen from the figs. 5 (c)
and 5 (d), the central minimum for $\tilde q$ is not observed when $\tilde
F^q$ is either positive or negative during most of the $\zeta$ region
covered. Thus the pattern in
$\sigma$ space depends closely on the $x, \zeta$ dependence of the photon
GPDs. The height of the maxima of both $q(x, \sigma,t) $ and $\tilde q(x,
\sigma, t)$ decrease with increasing $-t$ for a given $x$.             
\section{Conclusion}
In this work, we have calculated the GPDs of the photon when the momentum
transfer has both transverse and longitudinal components, in other words,
when both $\zeta$ and $\Delta^\perp$ are non-zero. We calculated both the
polarized and unpolarized GPDs in terms of the light-front wave functions
of the photon which, in turn, can be calculated in perturbation theory. The
helicity of the photon may flip in the process. In the present work we
considered only the non-helicity flip terms. We calculated them at leading
order in electromagnetic coupling $\alpha$ and zeroth order in strong
coupling $\alpha_s$. We also kept the subleading mass terms in the vertex.
The GPDs depend logarithmically on the scale. However, here we studied the $x,
\zeta$ and $t$ dependence at a fixed scale. By taking a Fourier transform
with respect to $\Delta^\perp$ we expressed the GPDs in the transverse
impact parameter space. As already observed for zero $\zeta$, the photon GPDs
show distinctive features compared to the proton GPDs in impact parameter
space. For non-zero $\zeta$ similar featured are observed. We introduced a
boost invariant longitudinal impact parameter $\sigma$ conjugate to $\zeta$.
Taking a FT with respect to $\zeta$ we expressed the GPDs in longitudinal
position space. They show an interesting pattern similar to the diffraction
pattern in optics. This was also observed for the GPDs of the proton as well
as in the DVCS amplitude for a dressed electron. We presented a comparative
analysis of the behaviour of the photon GPDs and that of a proton in $\sigma$
space and showed that the finite range of the $\zeta$ integration as well as
the $x,\zeta$ dependence of the photon GPDs determine the pattern in the
longitudinal position space.  
As is well known, the substructure of the photon is of relevance 
in high energy processes. Here we showed  that the photon GPDs can provide a
unique picture of the structure of the photon in 3D position space. Further
investigation would involve a study of the photon GPDs when the helicity is
flipped. It would be interesting to understand if the GPDs of the photon obey
the general properties like polynomiality and positivity. A violation of
positivity was seen in the transverse impact parameter space in \cite{us1}
and that needs to be investigated further. One has to remember also that a
complete picture of the GPDs in $\sigma$ space would need the contribution
from the region $x<\zeta$. The analysis presented here provides a
major step towards understanding these objects.

\section{Acknowledgments}
This work is supported by BRNS grant Sanction No. 2007/37/60/BRNS/2913 
dated 31.3.08, Govt. of India. We thank B. Pire for suggesting this topic
and for helpful discussions.     


\end{document}